\documentclass[10pt,conference]{IEEEtran}
\usepackage[tight,footnotesize]{subfigure}
\usepackage{booktabs, multicol, multirow}
\usepackage[acronym,nomain]{glossaries}
\usepackage[inline]{enumitem}
\usepackage[colorinlistoftodos,prependcaption,textsize=tiny]{todonotes}
\usepackage{algorithm}
\usepackage{algorithmicx}
\usepackage{algpseudocode}
\usepackage{amsmath,amsfonts,amssymb}
\usepackage{subfigure}
\usepackage{grffile}
\usepackage{tabularx}
\usepackage{colortbl}
\usepackage{hhline}
\usepackage[bookmarks=false]{hyperref}
\usepackage[flushleft]{threeparttable}
\usepackage{booktabs,caption,fixltx2e}
\usepackage{balance}

\renewcommand{\mathbf}[1]{{\boldsymbol #1}}
\hypersetup{
	colorlinks   = true,
	citecolor    = blue,
	linkcolor    = blue,
}
\usepackage{xcolor}
\usepackage[tikz]{bclogo}
\usepackage[framemethod=tikz]{mdframed}
\usepackage{lipsum}
\usepackage[many]{tcolorbox}
\usepackage{framed}
\colorlet{shadecolor}{blue!20}
\newcolumntype{g}{>{\columncolor{blue!25}}c}

\newcommand{\figref}[1]{Figure~\ref{fig:#1}}

\makeglossaries

\begin{document}

\title{Transfer Learning for Improving Model Predictions in Highly Configurable Software}

\author{
	\IEEEauthorblockN{Pooyan Jamshidi, Miguel Velez, Christian K{\"a}stner}
	\IEEEauthorblockA{
	Carnegie Mellon University, USA \\
	\{pjamshid,mvelezce,kaestner\}@cs.cmu.edu
	}

		\and
    \IEEEauthorblockN{Norbert Siegmund}
	\IEEEauthorblockA{Bauhaus-University Weimar, Germany \\
 	norbert.siegmund@uni-weimar.de
	}
		\and
	\IEEEauthorblockN{Prasad Kawthekar}
	\IEEEauthorblockA{Stanford University, USA \\
	pkawthek@stanford.edu
	}
}

\maketitle
\begin{abstract}

Modern software systems are built to be used in dynamic environments using configuration capabilities to adapt to changes and external uncertainties. In a self-adaptation context, we are often interested in reasoning about the performance of the systems under different configurations. Usually, we learn a black-box model based on real measurements to predict the performance of the system given a specific configuration. However, as modern systems become more complex, there are many configuration parameters that may interact and we end up learning an exponentially large configuration space. Naturally, this does not scale when relying on real measurements in the actual changing environment.
We propose a different solution: Instead of taking the measurements from the real system, we learn the model using samples from other sources, such as simulators that approximate performance of the real system at low cost. We define a cost model that transform the traditional view of model learning into a multi-objective problem that not only takes into account model accuracy but also measurements effort as well.
We evaluate our cost-aware transfer learning solution using real-world configurable software including (i) a robotic system, (ii) 3 different stream processing applications, and (iii) a NoSQL database system. The experimental results demonstrate that our approach can achieve (a) a high prediction accuracy, as well as (b) a high model reliability. % with only a few samples from the target environment.
%As learning method, we use Gaussian Process (GP) models to learn the mean estimations. %The benefit of GP models is that they provide a confidence interval for each estimation so that we are able to reason about the uncertainty of the model predictions.

\end{abstract}


%!TEX root = ../paper.tex
\section{Introduction}
\label{sec:introduction}

Most software systems today are configurable, which gives end users, developers, and administrators the chance to customize the system to achieve a different functionality or tune its performance. In such systems, hundreds or even thousands of configuration parameters can be tweaked, making the system highly configurable~\cite{influence}. The exponentially growing configuration space, complex interactions, and unknown constraints among configuration options make it difficult to understand the performance of the system. As a consequence, many users rely on default configurations or they change only individual options in an ad-hoc way.

In this work, we deal with the type of configurable systems that operate in dynamic and uncertain environments (\emph{e.g.}, robotic systems). Therefore, it is desirable to react to environmental changes by tuning the configuration of the system when we anticipate that the performance will drop to an undesirable level. To do so, we use black-box performance models that describe how configuration options and their interactions influence the performance of a system (\emph{e.g.}, execution time). Black-box performance models are meant to ease understanding, debugging, and optimization of configurable systems~\cite{influence}. For example, a reasoning algorithm may use the learned model in order to identify the best performing configuration for a robot when it goes from indoor to an outdoor environment. % (as a result the configuration options regarding localizations may be adapted).

\begin{figure}[t]
	\begin{center}
		\includegraphics[width=0.7\columnwidth]{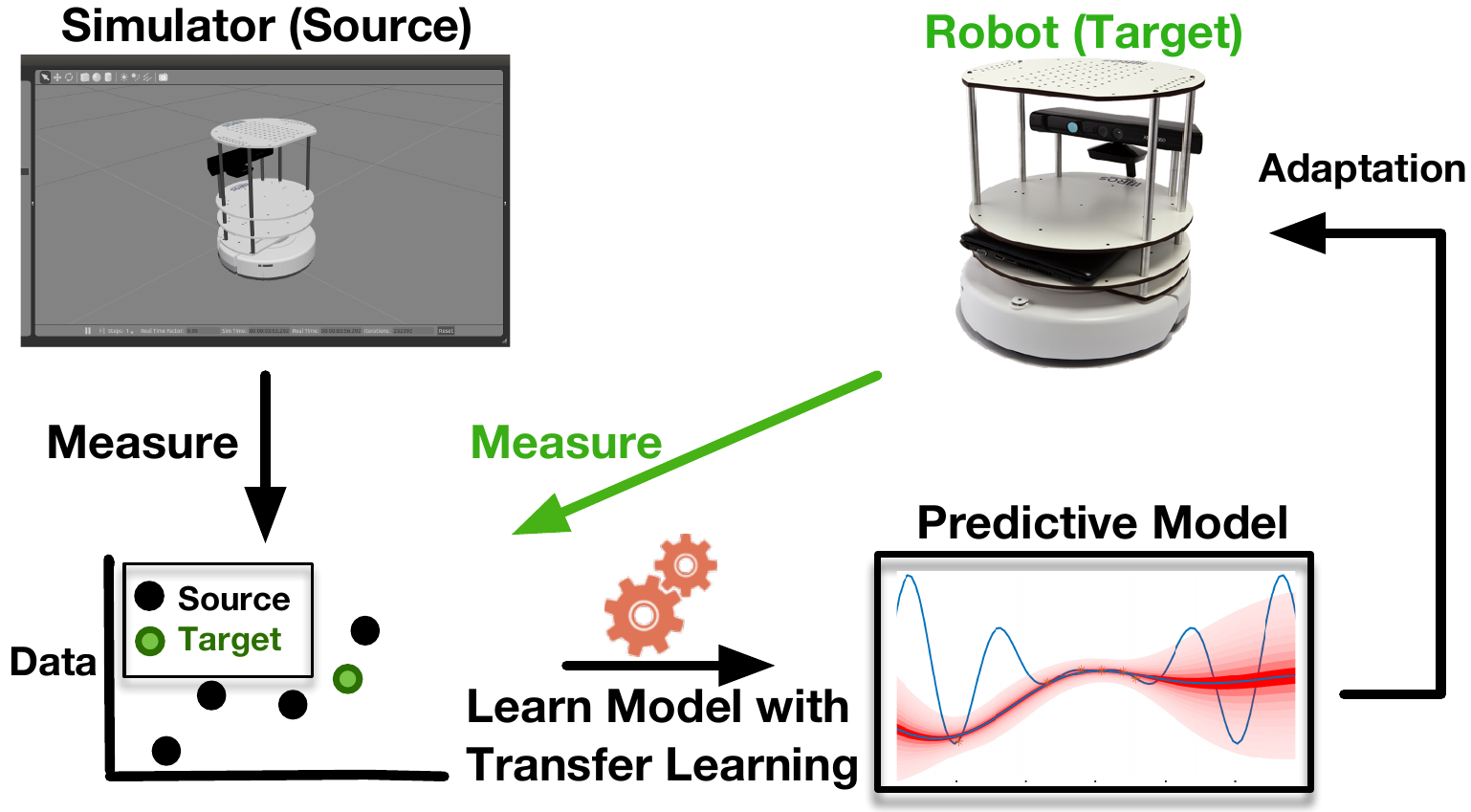}
		\caption{Transfer learning for performance model learning.}
		\label{fig:overview}
	\end{center}
\end{figure}

Typically, we learn a performance model for a given configurable system by measuring from a set of configurations selected by some sampling strategy. That is, we measure the performance of a given system multiple times in different configurations and learn how the configuration options and their interactions affect performance. However, such a way of learning from real systems, whether it is a robot or a software application, is a difficult task for several reasons: (i)~environmental changes (\emph{e.g.}, people wandering around robots), (ii)~high costs or risks of failure (\emph{e.g.}, a crashed robot), (iii)~the large amount of time required for measurements (\emph{e.g.}, we have to repeat the measurements several times to get a reliable value), and (iv)~changing system dynamics (\emph{e.g.}, robot motion). Moreover, it is often not possible to create potentially important scenarios in the real environment.

In this paper, as depicted in \figref{overview}, we propose a different solution: instead of taking the measurements from the real system, we reuse prior information (that we can get from other sources at a lower cost) in order to learn a performance model for the real system faster and cheaper. The concept of reusing information from other sources is the idea behind \emph{transfer learning}~\cite{torrey2009transfer,pan2010survey}. Similar to human beings that can learn from previous experience and transfer the learning to accomplish new tasks easier, quicker, and in a better way, in this work, we use other sources to provide cheaper samples for accelerating model learning. Instead of taking the measurements from the real system (we refer to as the \emph{target}), we measure the system performance using a proxy of the system (we refer to as the \emph{source}, \emph{e.g.}, a simulator). We then use a regression model that automatically learns the relationship between source and target to learn an accurate and reliable performance model using only a few samples taken from the real system, leading to much lower cost and faster learning period. 
We define a cost model that turns model learning into a multi-objective problem taking into account model accuracy as well as measurement cost. We demonstrate that our cost-aware transfer learning will enable accurate performance predictions by using only a few measurements of the target system. We evaluate our approach on (i) a robotic system, (ii) 3 stream processing applications, and (iii) a NoSQL database system. %We demonstrate that our approach enhances the model prediction accuracy with at least an order of magnitude comparing with the model that has been learned using only the measurements on the real system.

In summary, our contributions are the following:
\begin{itemize}
	\item A \emph{cost-aware transfer learning} method that learns accurate performance models for configurable software.
	\item An implementation, analysis and \emph{experimental evaluation} of our cost-aware transfer learning compared to a no transfer learning method.
	%\item \emph{Measurement datasets} worth several months of experimental time for the subject systems.
\end{itemize}

%In the rest, Section \ref{sec:background} overviews the problem and motivates the work via an example. Our transfer learning methodology is introduced in Section \ref{sec:approach} and then validated in Section \ref{sec:evaluation}. Section \ref{sec:related-work} positions the approach int he literature and Section \ref{sec:conclusions} concludes the paper.

%!TEX root = ../paper.tex
\section{Model Learning for Performance Reasoning in Highly Configurable Software}
\label{sec:background}

We motivate the model learning problem by reviewing specific challenges in the robotics domain. However, our method can be applied in other configurable systems as well. %Later, for the sake of evaluation, we will demonstrate the results with 3 different classes of system: (i) robotics, (ii) stream processing, and (iii) NoSQL database.

\subsection{A motivating example}
\label{sec:example}

The software embedded in a robot implements algorithms that enable the robot to accomplish the missions assigned to it, such as path planning~\cite{kawthekarsensitivity}. %The quality of the algorithms and their implementation thus determine the performance we observe from the robot.
Often, robotics software exposes many different configuration parameters that can be tuned and typically affect the robot's performance in accomplishing its missions.
For instance, we would like to tune the localization parameters of an autonomous robot while performing a navigation task as \emph{fast} and \emph{safe} as possible while consuming \emph{minimal energy} (corresponding to longer battery life).
%The configuration parameters may be hidden in the source code or in a configuration file. They may be different in terms of granularity, from categorical to real values~\cite{Rabkin:2011}.

Navigation is the most common activity that a robot does in order to get from point A to point B in an environment as efficient as possible without hitting obstacles, walls, or people. Let us concentrate on the \emph{localization} algorithm that enables navigation tasks and previous studies shown that configuration setting is influential to its performance \cite{kawthekarsensitivity}. Localization is the technique that enables the robot to find (i) its current position and (ii) its current orientation at any point in its navigation task~\cite{thrun2005probabilistic}.
Configurations that work well for one robot in one environment may not work well or at all in another environment for multiple reasons: %For example, localization configurations which work well in an outdoors environment may be ineffective indoors. This motivated us to consider changing the configuration parameters relevant to the localization of autonomous robots that typically operate in dynamic environments in order to improve the accuracy of the localization task that leads to improving the efficiency of its navigation.
%There are specific reasons why we need to change the configuration of robots at runtime:
(i) \textbf{The environment in which robots operate may change at runtime}. For instance, the robot may start in a dark environment and then reaches a much brighter area. As a result, the quality of localization may be affected as the range sensors for estimating the distance to the wall provide measurements with a higher error rate and the false readings increase then. %As a result, the robot may require more time to localize in the environment.
(ii) \textbf{The robot itself may move to an environment where the motion of the robot will change}. For example, when a wheeled robot moves to a slippery floor, then the motion dynamics of the robot will change and the robot cannot accelerate the speed with the same power level.
(iii) \textbf{The physics of the environment may change}. For example, imagine the situation where some items are put on a service robot while doing a mission. As a result, the robot has to sustain more weight and consumes more energy for doing the same mission. Therefore, it may be desirable to sacrifice the accuracy by adjusting localization parameters in order to perform the mission before running out of battery.
(iv) \textbf{A new mission may be assigned to the robot}. While a service robot is doing a mission, a new mission may be assigned, for example, a new delivery spot may be defined. In cases where the battery level is too low to finish the new mission, we would change to a configuration that sacrifices the localization accuracy in favor of energy usage.

In all of these situations, it is desirable to change the configuration regarding the localization of the robot, in an automated fashion. For doing so, we need a model to reason about the system performance especially when we face contradicting aspects that require a trade-off at runtime. %For instance, we may want to sacrifice performance for a lower battery usage for the service robot in order to finish the navigation task before running out of battery.
%As a results of these dynamic changes to the environment and the robot itself, we need to find a new configuration parameter that works well for the localization algorithm and as a result the robot will be able to accomplish the task \emph{safely} with \emph{minimum battery usage}.
We can learn a performance model empirically using observations which are taken from the robot performing missions in a real environment under different configurations. But, the observations in real-world are typically costly, error-prone, and sometimes infeasible. There are, however, several simulation platforms that provide a simulated environment, in which we can learn about the performance of the robot in different conditions given a specific configuration (\emph{e.g.}, Gazebo \cite{reckhaus2010overview}). Although the observations from a simulated environment are far less costly than the measurements taken from the real robot and impose no risks when a failure happens, they may not necessarily reflect real behavior, since not every physical aspect can be accurately modeled and simulated. Fortunately, simulations are often highly \emph{correlated} with the real behavior. The key insight behind our approach is to learn the robot performance model by using only a few expensive observations on a real system, but several cheap samples from all sources. %We can then exploit the knowledge we gain from the relationship of simulation samples and the few ones on real system to train a performance model that is more accurate at less costly comparing with the situation where we rely only on observations from real systems.

%The localization algorithm in robotics platforms (\emph{e.g.}, ROS) comprises of hundreds parameters and the configuration space is simply too large to learn a reliable model relying on sampling the real robot platform. There are, however, several simulation platforms that provide a simulated environment in which we can learn about the performance of the robot in different conditions given a specific configuration (\emph{e.g.}, Gazebo). Now the key challenge is how to learn a model mainly from the measurements from simulation (which are far less costly than the measurements on the real robot platform and does not have the risks mentioned earlier).

\subsection{Challenges}

Although we can take relatively cheap samples from simulation, it is impractical and naive to exhaustively run simulation environments for all possible configurations:
\begin{itemize}
\item \textbf{Exponentially growing space}. The configuration space of just 20 parameters with binary options for each comprises of $2^{20}\approx1m$ possible configurations. Even for this small configuration space, if we spend just one minute for collecting each sample, it will take about 2 years to perform an exhaustive sampling. Therefore, we can sample only a subset of this space, and we need to do the sampling purposefully.
\item \textbf{Negative transfer}. Not all samples from a simulator reflect the real behavior and, as a result, they would not be useful for learning a performance model. More specifically, the measurements from simulators typically contain noise and for some configurations, the data may not have any relationship with the data we may observe on the real system. Therefore, if we learn a model based on these data, it becomes inaccurate and far from real system behavior. Also, any reasoning based on the model predictions becomes misleading or ineffective.
\item \textbf{Limited budget}. Often, different sources of data exist (\emph{e.g.}, different simulators, different versions of a system) that we can learn from and each may impose a different cost. Often, we are given a limited budget that we can spend for either source and target sample measurements.
%Moreover, too many training samples cause long training time. Since the learned model may be used in some time constrained environments (\emph{e.g.}, runtime decision making in a feedback loop for robots), it is important to select the sources purposefully.
\end{itemize}

\subsection{Problem formulation: Black-box model learning}
\label{sec:problem}

In order to introduce the concepts in our approach concisely, we define the model learning problem using mathematical notations. Let $X_i$ indicate the $i$-th configuration parameter, which ranges in a finite domain $Dom(X_i)$. In general, $X_i$ may either indicate (i) an integer variable (\emph{e.g.}, the \emph{number of iterative refinements} in a localization algorithm) or (ii) a categorical variable (\emph{e.g.}, \emph{sensor names} or binary options (\emph{e.g.}, \emph{local vs global localization method}). The configuration space is mathematically a Cartesian product of all of the domains of the parameters of interest $\mathbb{X}=Dom(X_1) \times \dots \times Dom(X_d)$. %A configuration $\mathbf{x}$ then resides in the configuration space $\mathbf{x} \in \mathbb{X}$.

A black-box response function $f:\mathbb{X}\rightarrow\mathbb{R}$ is used to build a performance model given some observations of the system performance under different settings. In practice, though, the observation data may contain noise, \emph{i.e.},~$y_i=f(\mathbf{x}_i)+\epsilon_i,\mathbf{x}_i \in \mathbb{X}$ where $\epsilon_i \sim \mathcal{N}(0,\sigma_i)$ and we only partially know the response function through the observations $\mathcal{D}=\{(\mathbf{x}_i,y_i)\}_{i=1}^d, |\mathcal{D}|\ll|\mathbb{X}|$.
In other words, a response function is simply a mapping from the configuration space to a measurable performance metric that produces interval-scaled data (here we assume it produces real numbers). Note that we can learn a model for all measurable attributes (including accuracy, safety if suitably operationalized), but here we mostly train predictive models on performance attributes (\emph{e.g.}, response time, throughput, CPU utilization).

Our main goal is to learn a reliable regression model, $\hat{f}(\cdot)$, that can predict the performance of the system, $f(\cdot)$, given a limited number of observations $\mathcal{D}$. More specifically, we aim to minimize the \emph{prediction error} over the configuration space:
\begin{equation} \label{eq:objective}
\arg \min _{\mathbf{x}\in\mathbb{X}} pe=|\hat{f}(\mathbf{x})-f(\mathbf{x})|
\end{equation}

In order to solve the problem above, we assume $f(\cdot)$ is reasonably smooth, but otherwise little is known about the response function. Intuitively, we expect that for ``near-by'' input points $\mathbf{x}$ and $\mathbf{x}'$ their corresponding output points $y$ and $y'$ to be ``near-by'' as well. %This allows us to build a model using black-box models that can predict unobserved values.

\subsection{State-of-the-art}
In literature, model learning has been approached from two standpoints: (i) sampling strategies and (ii) learning methods.

\subsubsection{Sampling}
Random sampling has been used to collect unbiased observations in computer-based experiments. However, random sampling may require a large number of samples to build an accurate model \cite{influence}.
More intelligent sampling strategies (such as Box-Behnken and Plackett-Burman) have been developed in the statistics and machine learning communities, under the umbrella of experimental design, to ensure certain statistical properties~\cite{montgomery2008design}. The aim of these different experimental designs is to ensure that we gain a high level of information from sparse sampling (partial design) in high dimensional spaces. Relevant to the software engineering community, several approaches tried different designs for highly configurable software~\cite{guo2013variability} and some even consider cost as an explicit factor to determine optimal sampling \cite{sarkar2015cost}.

For finding optimal configurations, researchers have tried novel ways of sampling with a feedback embedded inside the process where new samples are derived based on information gained from the previous set of samples. A recent solution \cite{bodin2016integrating} uses active learning based on a random forest to find a good design.
Recursive Random Sampling (RRS)~\cite{ye2003recursive} integrates a restarting mechanism into the random sampling to achieve high search efficiency. Smart Hill Climbing (SHC)~\cite{xi2004smart} integrates importance sampling with Latin Hypercube Design (LHD). %SHC estimates the local regression at each potential region, then it searches toward the steepest descent direction. 
An approach based on direct search~\cite{Zheng2007} forms a simplex in the configuration space, and iteratively updates the simplex through a number of operations to guide the sample generation. Quick Optimization via Guessing (QOG)~\cite{osogami2007optimizing} speeds up the optimization process exploiting some heuristics to filter out sub-optimal configurations. Recently, transfer learning has been explored for configuration optimizations by exploiting the dependencies between configurations parameters \cite{chen2009experience} and measurements for previous versions of big data systems using an approach called TL4CO in DevOps \cite{artavc2017dice}.  

\subsubsection{Learning} 
Also, standard machine-learning techniques, such as support-vector machines, decision trees, and evolutionary algorithms have been tried~\cite{jamshidi2016bo4co,yigitbasi2013towards}. These approaches trade simplicity and understandability of the learned models for predictive power. For example, some recent work~\cite{zhang2015performance} exploited a characteristic of the response surface of the configurable software to learn Fourier sparse functions by only a small sample size. Another approach \cite{influence} also exploited this fact, but iteratively construct a regression model representing performance influences in an active learning process.

\subsubsection{Positioning in the self-adaptive community}
Performance reasoning is a key activity for decision making at runtime. Time series techniques~\cite{ehlers2011self} shown to be effective in predicting response time and uncovering performance anomalies ahead of time. FUSION~\cite{esfahani2013learning,elkhodary2010fusion} exploited inter-feature relationships (\emph{e.g.}, feature dependencies) to reduce the dimensions of configuration space, making runtime performance reasoning feasible. Different classification models have been evaluated for the purpose of time series predictions in \cite{anaya2014prediction}.
Performance predictions also have been applied for resource allocations at runtime~\cite{huber2017model,jamshidi2016managing}.
Note that the approaches above are referred to as black-box models. However, another category of models known as white-box is built early in the life cycle, by studying the underlying architecture of the system \cite{gomaa2007model,happe2011facilitating} using Queuing networks, Petri Nets, and Stochastic Process Algebras~\cite{balsamo2004model}.
Performance prediction and reasoning have also been used extensively in other communities such as component-based~\cite{becker2006performance} and control theory~\cite{filieri2015automated}.

\subsubsection{Novelty}
Previous work attempted to improve the prediction power of the model by exploiting the information that has been gained from the target system either by improving the sampling process or by adopting a learning method.
Our approach is orthogonal to both sampling and learning, proposing a new way to enhance model accuracy by exploiting the knowledge we can gain from other \emph{relevant} and possibly \emph{cheaper} sources to accelerate the learning of a performance model through transfer learning.
%Transfer learning attempts to transfer knowledge learned in one or more source tasks and leverage it to improve learning in a related target task \cite{torrey2009transfer}. In this context, source tasks are performance models of the system under different environmental conditions: (i)~different workload, (ii)~different deployment infrastructure or (iii)~different versions of the system. The performance of the system varies when one of these environmental conditions changes, however, as we will show later in our experimental results, the performance responses are correlated. This correlation is an indicator that the source and target are related and there is a potential to learn across the environments.

Transfer learning has been applied previously for regression and classification problems in machine learning \cite{pan2010survey}, in software engineering for defect predictions \cite{krishna2016too,nam2015heterogeneous,nam2013transfer} and effort estimation \cite{kocaguneli2015transfer} and in systems for configuration optimization \cite{chen2009experience,artavc2017dice}. However, in this paper, we enable a generic form of transfer learning for the purpose of sensitivity analysis over the entire configuration space. Our approach can enable (i) performance debugging, (ii) performance tuning, (iii) design-time evolution, or (iv) runtime adaptation in the target environment by exploiting any source of knowledge from the source environment including (i)~different workloads, (ii)~different deployments, (iii)~different versions of the system, or (iv)~different environmental conditions. The performance of the system varies when one or more of these changes happen, however, as we will show later in our experimental results, the performance responses are correlated. This correlation is an indicator that the source and target are related and there is a potential to learn across the environments. To the best of our knowledge, our approach is the first attempt towards learning performance models using cost-aware transfer learning for configurable software.
%!TEX root = ../paper.tex
\section{Cost-Aware Transfer Learning}
\label{sec:approach}

In this section, we explain our \emph{cost-aware transfer learning} solution to improve learning an accurate performance model.

\subsection{Solution overview}

\begin{figure}[t]
	\begin{center}
		\includegraphics[width=\columnwidth]{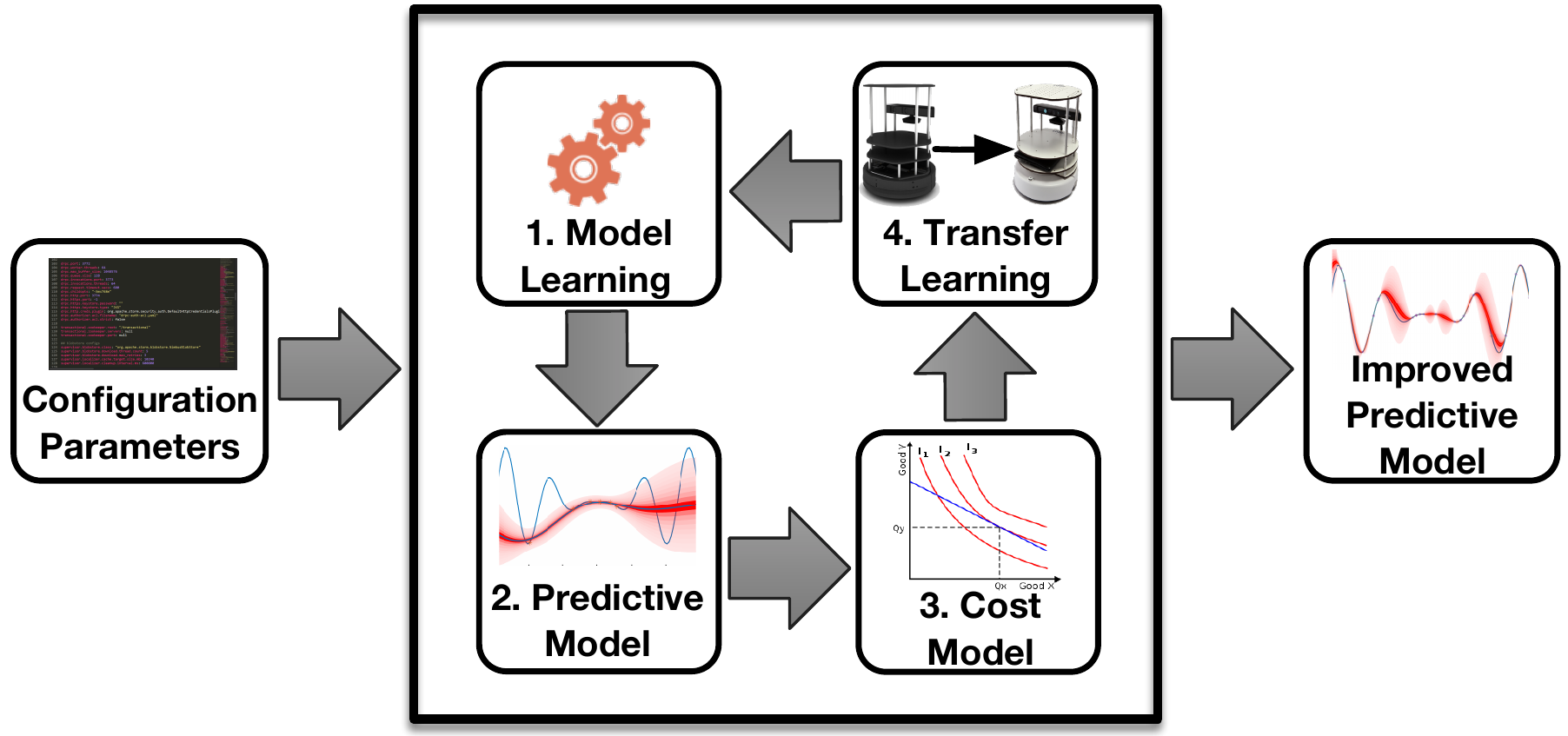}
		\caption{Overview of cost-aware transfer learning methodology.} %The learning process is based on a small subset of the overall configuration space that samples are taken from different sources and target.
		%Our methodology enables appropriate selection of samples from sources and target that gives the highest accuracy while satisfying budget.} %Subsequently, the predictive model can predict performance of an unseen configuration depending on its parameters.}
		\label{fig:prediction-tl}
	\end{center}
\end{figure}

An overview of our model learning methodology is depicted in Figure~\ref{fig:prediction-tl}. We use the observations that we can obtain cheaply, with the cost of $c_s$ for each sample, from an alternative measurable response function, $g(\cdot)$, that yields different but related response to the target response, $\mathcal{D}_s=\{(\mathbf{x}_s,y_s)|y_s=g(\mathbf{x}_s)+\epsilon_s, \mathbf{x}_s\in \mathbb{X}\}$, and transfer these observations to learn a performance model for the real system using only few observations from that system, $\mathcal{D}_t=\{(\mathbf{x}_t,y_t)|y_t=f(\mathbf{x}_t)+\epsilon_t, \mathbf{x}_t\in \mathbb{X}\}$.
The cost of obtaining each observation on the target system is $c_t$, and we assume that $c_s\ll c_t$ and that the source response function, $g(\cdot)$, is related (correlated) to the target function, and this relatedness contributes to the learning of the target function, using only sparse samples from the real system, \emph{i.e.}, $|\mathcal{D}_t|\ll |\mathcal{D}_s|$. Intuitively, rather than starting from scratch, we transfer the learned knowledge from the data we have collected using cheap methods such as simulators or previous system versions, to learn a more accurate model about the target configurable system and use this model to reason about its performance at runtime.
%The relatedness of the source and target response functions may appear in different forms. Figure~\ref{fig:fg} shows possible relationships in terms of (i) translation, (ii) noise, and (iii) discrepancies in boundaries.

\begin{figure}[t]
	\begin{center}
		\includegraphics[width=\columnwidth]{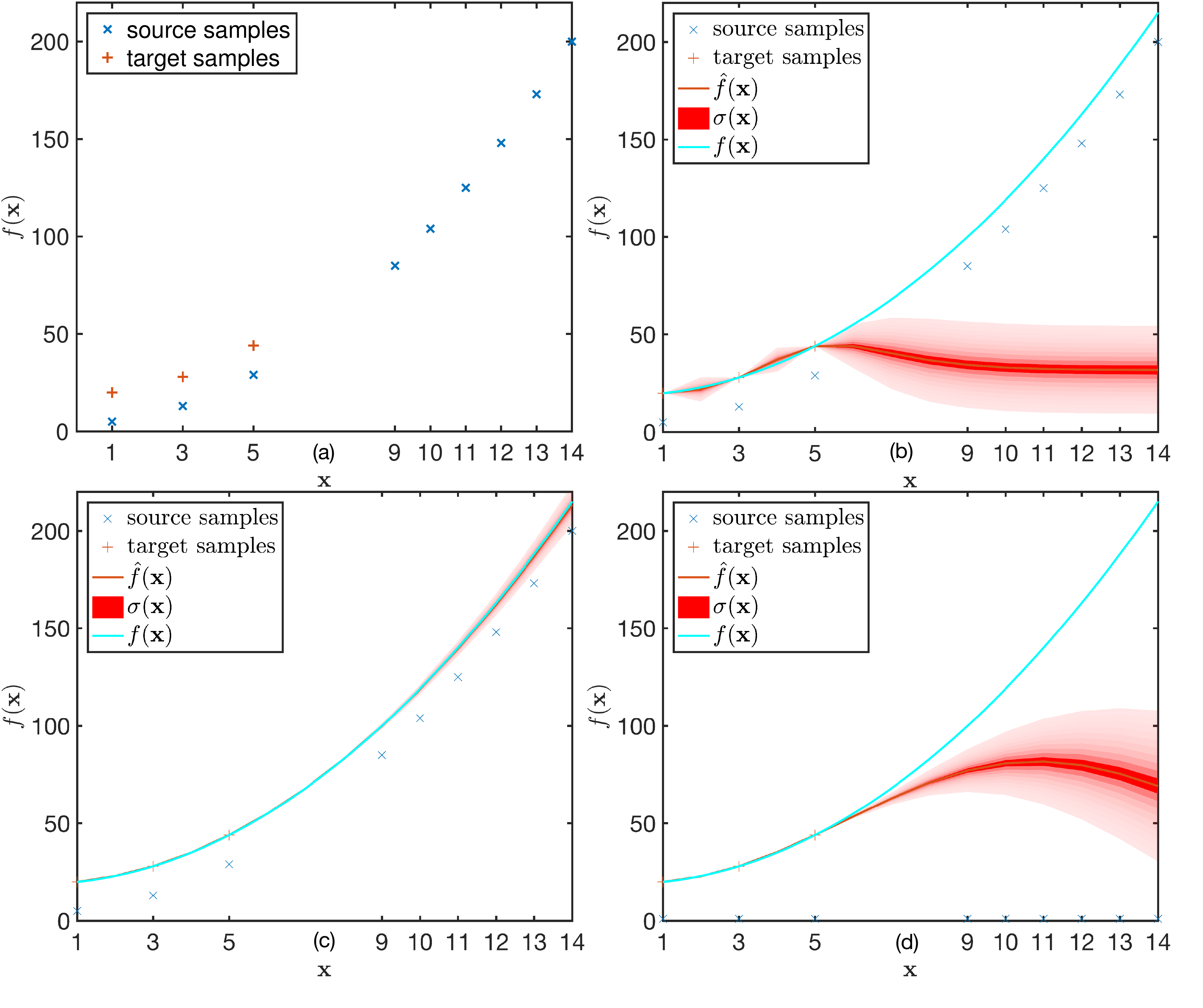}
		\caption{(a) 9 samples have been taken from a source (respectively 3 from a target); (b) [without transfer learning] a regression model is constructed using only the target samples. (c) [with transfer learning] another regression model is constructed using the target and the source samples. The new model is based on only 3 target measurements while providing accurate predictions. (d) [negative transfer] a regression model is constructed with a transfer from a misleading response function and therefore the model prediction is inaccurate.}
		\label{fig:example-synthetic}
	\end{center}
\end{figure}

In Figure~\ref{fig:example-synthetic}, we demonstrate the idea of transfer learning with a synthetic example. Only three observations are taken from a synthetic target function, $f(\cdot)$, and we try to learn a model, $\hat{f}(\mathbf{x})$, which provides predictions for unobserved response values at some input locations. Figure \ref{fig:example-synthetic}(b) depicts the predictions that are provided by the model trained using only the 3 samples without considering the transfer learning from any source. The predictions are accurate around the three observations, while highly inaccurate elsewhere. Figure \ref{fig:example-synthetic}(c) depicts the model trained over the 3 observations and 9 other observations from a different but related source. The predictions, thanks to transfer learning, are here closely following the true function with a narrow confidence interval (\emph{i.e.}, high confidence in predictions provided by the model). Interestingly, the confidence interval around points $\mathbf{x}=2,4$ in the model with transfer learning have disappeared, demonstrating more confident predictions with transfer learning.

Given a target environment, the effectiveness of any transfer learning depends on the source environment and how it is related to the target \cite{torrey2009transfer}. If the relationship is strong and the transfer learning method can take advantage of it, the performance in the target predictions can significantly improve via transfer. However, if the source response is not sufficiently related or if the transfer method does not exploit the relationship, the performance may not improve or even may decrease. We show this case via the synthetic example in Figure \ref{fig:example-synthetic}(d) where the prediction model uses 9 samples from a misleading function that is unrelated to the target function (the response remains constant as we increase $\mathbf{x}$); as a consequence, the predictions are very inaccurate and do not follow the growing trend of the true function. %Also the confidence interval is sharply increasing as we go further away from the the observations.

%\begin{figure}[h]
%	\begin{center}
%		\includegraphics[width=\columnwidth]{figures/task_relatedness}
%		\caption{Two strategies for sampling sources: (a) choose a source response from which to transfer, (b) use the knowledge of how related the sources are to the target.}
%		\label{fig:task_relatedness}
%	\end{center}
%\end{figure}

For the choice of the number of samples from the source and target, we use a simple \emph{cost model} (cf. Figure \ref{fig:prediction-tl}) that is based on the number of source and target samples as follows:
\begin{align} \label{eq:cost-model}
\mathcal{C}(\mathcal{D}_s,\mathcal{D}_t)&=c_s\cdot |\mathcal{D}_s|+c_t \cdot |\mathcal{D}_t|+\mathcal{C}_{Tr}(|\mathcal{D}_s|,|\mathcal{D}_t|), \\
\mathcal{C}(\mathcal{D}_s,\mathcal{D}_t)&\le\mathcal{C}_{max},  \label{eq:cost-constraint}
\end{align}
where $\mathcal{C}_{max}$ is the experimental budget and $\mathcal{C}_{Tr}(|\mathcal{D}_s|,|\mathcal{D}_t|)$ is the cost of model training, which is a function of source and target sample sizes.
Note that this cost model makes the problem we have formulated in Eq. \eqref{eq:objective} into a multi-objective problem, where not only \emph{accuracy} is considered, but also \emph{cost} can be considered in the transfer learning process.

%\begin{figure*}[t]
%	\begin{center}
%		\includegraphics[width=\textwidth]{figures/fg}
%		\caption{Relationships between source and target response functions: (a) source is highly correlated with the target function but shifted or linearly skewed, (b) contain noise, (c) have discrepancies towards the boundaries, or (d) uncorrelated.}
%		\label{fig:fg}
%	\end{center}
%\end{figure*}

%\begin{figure}
%	\begin{center}
%		\includegraphics[width=0.7\columnwidth]{figures/pareto-synthetic-annotated}
%		\caption{Indifference curves representing combinations of $(|\mathcal{D}_t|,|\mathcal{D}_t|)$ associated with equal levels of prediction errors.}
%		\label{fig:pareto-synthetic}
%	\end{center}
%\end{figure}

\subsection{Model learning: Technical details}
\label{sec:gp}

We use Gaussian Process (GP) models to learn a reliable performance model $\hat{f}(\cdot)$ that can predict unobserved response values. The main motivation to choose GP here is that it offers a framework in which performance reasoning can be done using mean estimates as well as a confidence interval for each estimation. In other words, where the model is confident with its estimation, it provides a small confidence interval; on the other hand, where it is uncertain about its estimations it gives a large confidence interval meaning that the estimations should be used with precautions. The other reason is that all the computations are based on linear algebra which is cheap to compute. This is especially useful in the domain of self-adaptive systems where automated performance reasoning in the feedback loop is typically time constrained and should be robust against any sort of uncertainty including prediction errors. Also, for building GP models, we do not need to know any internal details about the system; the learning process can be applied in a black-box fashion using the sampled performance measurements. In the GP framework, it is also possible to incorporate domain knowledge as prior, if available, which can enhance the model accuracy \cite{jamshidi2016bo4co}.

In order to describe the technical details of our transfer learning methodology, let us briefly describe an overview of GP model regression; a more detailed description can be found elsewhere \cite{gpml}. GP models assume that the function $\hat{f}(\mathbf{x})$ can be interpreted as a probability distribution over functions:
\begin{equation} \label{eq:gp}
\mathbf{y}=\hat{f}(\mathbf{x}) \sim \mathcal{GP} (\mu(\mathbf{x}), k(\mathbf{x}, \mathbf{x}')),
\end{equation}
where $\mu:\mathbb{X}\rightarrow\mathbb{R}$ is the mean function and $k:\mathbb{X}\times\mathbb{X}\rightarrow\mathbb{R}$ is the covariance function (kernel function) which describes the relationship between response values, $\mathbf{y}$, according to the \emph{distance} of the input values $\mathbf{x},\mathbf{x}'$.
%Let us assume $\mathcal{D}=\{(\mathbf{x}_{i},y_{i}) | y_i:=f(\mathbf{x}_i)+\epsilon_i,i=1,\cdots,t\}$ be the collection of $t$ observations.
The mean and variance of the GP model predictions can be derived analytically \cite{gpml}:
\begin{align}
\mu_{t}(\mathbf{x})&=\mu(\mathbf{x})+\mathbf{k}(\mathbf{x})^\intercal (\mathbf{K}+\sigma^2\mathbf{I})^{-1} (\mathbf{y}-\boldsymbol{\mu}), \label{eq:gp-surrogate-mean}\\
\sigma_t^2(\mathbf{x})&=k(\mathbf{x},\mathbf{x})+\sigma^2\mathbf{I} - \mathbf{k}(\mathbf{x})^\intercal (\mathbf{K}+\sigma^2\mathbf{I})^{-1} \mathbf{k}(\mathbf{x}), \label{eq:gp-surrogate-sigma}
\end{align}
where $\mathbf{k}(\mathbf{x})^\intercal=[k(\mathbf{x},\mathbf{x}_1) \quad k(\mathbf{x},\mathbf{x}_2) \quad \dots \quad k(\mathbf{x},\mathbf{x}_t)]$, $\mathbf{I}$ is identity matrix and
\begin{equation} \label{eq:covariance}
\mathbf{K}:=
\begin{bmatrix}
k(\mathbf{x}_1,\mathbf{x}_1)  &  \dots & k(\mathbf{x}_1,\mathbf{x}_t)   \\
\vdots  & \ddots &  \vdots \\
k(\mathbf{x}_t,\mathbf{x}_1)  &  \dots & k(\mathbf{x}_t,\mathbf{x}_t)
\end{bmatrix}
\end{equation}

%In order to intuitively demonstrate the GP model predictions, we illustrate it through a one dimensional model in Figure \ref{fig:example-gp}. In the figure, the true function we try to predict using GP model, the mean estimation and the confidence interval at each point, cf. \eqref{eq:gp-surrogate-mean} and \eqref{eq:gp-surrogate-sigma}, and the observations are shown. Parts of the response function that is well explored (\emph{i.e.}, we have observations for) has small confidence interval, \emph{i.e.}, the variance of the predictions around those points are low. On the other hand, the parts of the space that are now well explored, the model is not confident enough with respect to the mean estimates and therefore the variances are high.

%\begin{figure}
%	\begin{center}
%		\includegraphics[width=0.73\columnwidth]{figures/gp-example2}
%			\caption{An example of one-dimensional GP model. Stars indicate the location of the data. The GP estimations, shown in red, runs along the means of the normally distributed confidence intervals shown in red shades. } %The solid green line is the mean prediction given the data, and the shaded grey area shows the confidence interval.}
%		\label{fig:example-gp}
%	\end{center}
%\end{figure}
GP models have shown to be effective for performance predictions in data scarce domains \cite{jamshidi2016bo4co}. However, as we have demonstrated in Figure \ref{fig:example-synthetic}, it may become inaccurate when the samples do not cover the space uniformly. For highly configurable systems, we require a large number of observations to cover the space uniformly, making GP models ineffective in such situations.

\subsection{Model prediction using transfer learning}
\label{sec:tl}

In transfer learning, the key question is how to make accurate predictions for the target environment using observations from other sources, $\mathcal{D}_s$. We need a measure of relatedness not only between input configurations but between the sources as well. The relationships between input configurations was captured in the GP models using the covariance matrix that was defined based on the kernel function in Eq. \eqref{eq:covariance}. More specifically, a kernel is a function that computes a dot product (a measure of ``similarity'') between two input configurations. So, the kernel helps to get accurate predictions for similar configurations. We now need to exploit the relationship between the source and target functions, $g,f$, using the current observations $\mathcal{D}_s,\mathcal{D}_t$ to build the predictive model $\hat{f}$.
To capture the relationship, we define the following kernel function:
\begin{align}
\label{eq:mt-kernel}
k(f,g,\mathbf{x},\mathbf{x}')=k_t(f,g)\times k_{xx}(\mathbf{x},\mathbf{x}'),
\end{align}
where the kernels $k_t$ represent the correlation between source and target function, while $k_{xx}$ is the covariance function for inputs. Typically, $k_{xx}$ is parameterized and its parameters are learned by maximizing the marginal likelihood of the model given the observations from source and target $\mathcal{D}=\mathcal{D}_s\cup\mathcal{D}_t$. Note that the process of maximizing the marginal likelihood is a standard method \cite{gpml}.
After learning the parameters of $k_{xx}$, we construct the covariance matrix exactly the same way as in Eq. \ref{eq:covariance} and derive the mean and variance of predictions using Eq. \eqref{eq:gp-surrogate-mean}, \eqref{eq:gp-surrogate-sigma} with the new $\mathbf{K}$. The main essence of transfer learning is, therefore, the kernel that captures the source and target relationship and provides more accurate predictions using the additional knowledge we can gain via the relationship between source and target.

%Note that $k_t$ depends only on the labels ${j}$ pointing to the versions for which we have some observations, and $k_{xx}$ depends only on the configuration $\mathbf{x}$.
%More specifically, we normalize $k_t$ to be a correlation matrix, so that it has ones along its diagonal. Then the extent of relatedness between any source and target is measured by their correlation. This matrix can be extended accordingly to accommodate for multiple sources, but here we assume we have one source.

%Figure \ref{fig:example-synthetic}(c) illustrates the predictions provided by an updated GP model using this idea. Further observations are available on another related source, whereas we only have few samples on the target function. Merely using these few samples would result in poor (uninformative) predictions. Using the correlation with the other two functions enables the GP model to provide more accurate predictions with a much higher confidence throughout estimations.

\subsection{Transfer learning in a self-adaptation loop}

Now that we have described the idea of transfer learning for providing more accurate predictions, the question is whether such an idea can be applied at runtime and how the self-adaptive systems can benefit from it. More specifically, we now describe the idea of model learning and transfer learning in the context of self-optimization, where the system adapts its configuration to meet performance requirements at runtime. The difference to traditional configurable systems is that we learn the performance model online in a feedback loop under time and resource constraints. Such performance reasoning is done more frequently for self-adaptation purposes. %For example, such models can be applied to change the configuration of a robot to consume less energy (\emph{e.g.}, swap the current localization with a less demanding one) to finish its mission in the situation where under the current configuration it will run out of the battery.

\begin{figure}[t]
	\begin{center}
		\includegraphics[width=\columnwidth]{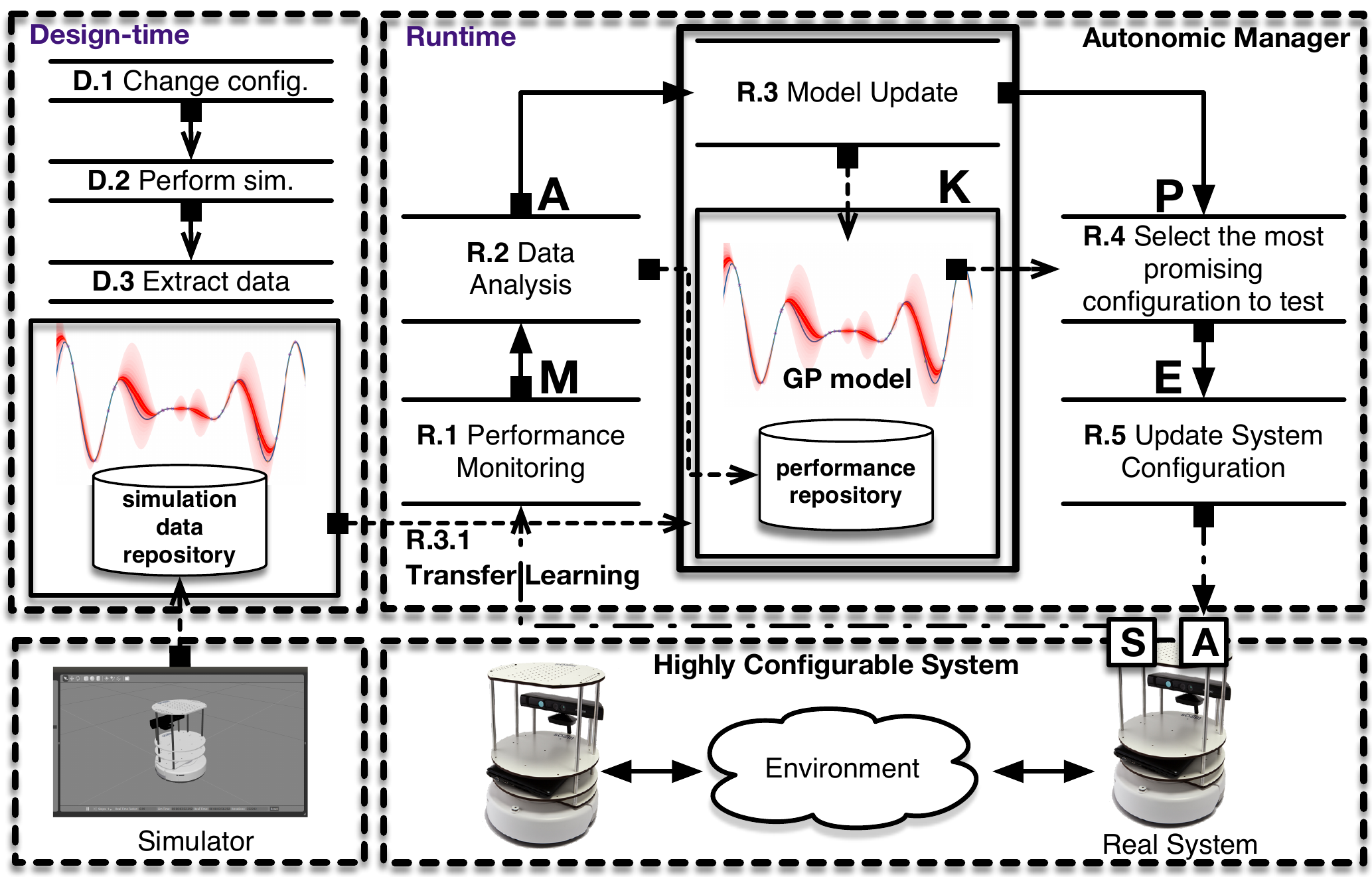}
		\caption{Integrating transfer learning with MAPE-K loop, where knowledge update is realized with transfer learning.}
		\label{fig:mape-ke}
	\end{center}
\end{figure}

An overview of a self-optimization solution is depicted in Figure \ref{fig:mape-ke} following the MAPE-K framework~\cite{de2013software,kephart2003vision}. We consider the regression model that we learn via transfer learning in this work as the \underline{K}nowledge component of the MAPE-K that acts as an interface to which other components can query the performance. %We use transfer learning to make the knowledge more accurate using observations that are taken from a simulator or any other cheap sources. 
For deciding how many observations and from what source to transfer, we use the cost model that we have introduced earlier. At runtime, the managed system is \underline{M}onitored by pulling the end-to-end performance metrics (\emph{e.g.}, latency, throughput) from the corresponding sensors. Then, the retrieved performance data needs to be \underline{A}nalysed. Next, the model needs to be updated taking into account the new performance observations. Having updated the model, a new configuration may be \underline{P}lanned to replace the current configuration. Finally, the new configuration will be enacted by \underline{E}xecuting platform specific operations. This enables model-based knowledge evolution using machine learning~\cite{jamshidi2016managing}. 

The underlying model can be updated not only when a new observation is available but also by transferring the learning from other related sources. So at each adaptation cycle, we can update our belief about the correct response given data from the managed system and other related sources. This can be particularly useful when the adaptive systems need to make a decision when the internal knowledge utilized in the MAPE-K loop is not accurate enough, \emph{e.g.}, in early stages of learning when the system needs to react to environmental changes the internal knowledge is not accurate and as a result the adaptation decisions are sub-optimal \cite{jamshidi2016fuzzy}. In this situation, even the measurements from a noisy source (\emph{e.g.}, simulator) could be beneficial in order to boost the initial performance achievable using only the transferred knowledge, before any further learning is done. Also, another challenge that has been reported in the past is the slow convergent of the learning process \cite{jamshidi2016fuzzy}, in this situation, transfer learning can accelerate the learning process in the target environment given the transferred knowledge compared to the amount of time to learn it from scratch. %In this situation, the transfer can also be done from multiple sources with different degrees of relatedness to the target (\emph{e.g.}, different simulators with different level of fidelity).

%Note that in the adaptation process a configuration may fail, cf. Section \ref{sec:problem}. In this situation, we need an approach that learns this over time and avoid trying the configuration that may most probably will fail in a certain situation. More specifically, we can learn a classifier that approximates a constrain function, $c(\cdot)$, that has been discussed in Section \ref{sec:problem}. Therefore, at runtime, when the classifier predict that with a high chance the configuration will fail, we may want to avoid changing the current configuration to the new one but to the one that may provide a slightly lower performance but with a less chance of failure.

%!TEX root = ../paper.tex

\section{Experimental results}
\label{sec:evaluation}

We evaluate the \emph{effectiveness} and \emph{applicability} of our transfer learning approach for learning models for highly-configurable systems, in particular, compared to conventional non-transfer learning. Specifically, we aim to answer the following three research questions:

{\noindent \em \textbf{RQ1}: How much does transfer learning improve the prediction accuracy?} %What is the difference in terms of range of prediction errors we get with our transfer learning approach comparing with no transfer learning?}

{\noindent \em \textbf{RQ2}: What are the trade-offs between learning with different numbers of samples from source and target in terms of prediction accuracy?}

{\noindent \em \textbf{RQ3}: Is our transfer learning approach applicable in the context of self-adaptive systems in terms of model training and evaluation time?}

We will proceed as follows. First, we will return to our motivating example of an autonomous service robot to explore
the benefits and limitations of transfer learning for this single case. Subsequently, we
explore the research questions quantitatively, performing experiments on 5 different configurable systems.
We have implemented our cost-aware transfer learning approach in Matlab 2016b and, depending on each experiment, we have developed different scripts for data collections and automated configuration changes. The source code and data are available in an online appendix~\cite{onlineappendix}.

%We first start the evaluation with a robotics system where the configuration space that we have considered is 2-dimensional. We present this as a case study in which we explore the feasibility and preparation of it. We then systematically approach the evaluation by controlling the sampling from both source and target configurations. We have conducted three systematic experiments to (i) evaluate the prediction error range comparing with the no transfer learning (RQ1) and (ii) to construct a Pareto front to discuss the trade-off between measurement costs and model accuracy (RQ2). In a third experiment, we assess the feasibility of our approach for self-adaptive systems discussing the training and testing efforts for model constructions and model update at runtime (RQ3).

\subsection{Case study: Robotics software}
\label{sec:case-study}
We start by demonstrating our approach with a small case
study of service robots, which we already motivated in Section~\ref{sec:example}.
Our long-term goal is to allow the service robots to adapt
effectively to (possibly unanticipated) changes in the environment, in
its goals, or in other parts of the system, among others, by reconfiguring
parameters. Specifically, we work with the CoBot robotic platform~\cite{veloso2015cobots}.

To enable a more focused analysis and presentation, we focus
on a single but important subsystem and two parameters:
The particle-filter-based autonomous localization component of the CoBot software~\cite{biswas2013localization}
determines the current location of the robot, which is essential for navigation.
Among the many parameters of the component, we analyze two that strongly
influence the accuracy of the localization and required computation effort,
(1)~the number of particle estimates and (2)~the number of gradient descent refinement
steps applied to each of these particles when we receive a new update from sensors.
At runtime, the robot could decide to reconfigure its localization
component to trade off location accuracy with energy consumption,
based on the model we learn.

\begin{figure}[t]
	\begin{center}
		\includegraphics[width=\columnwidth]{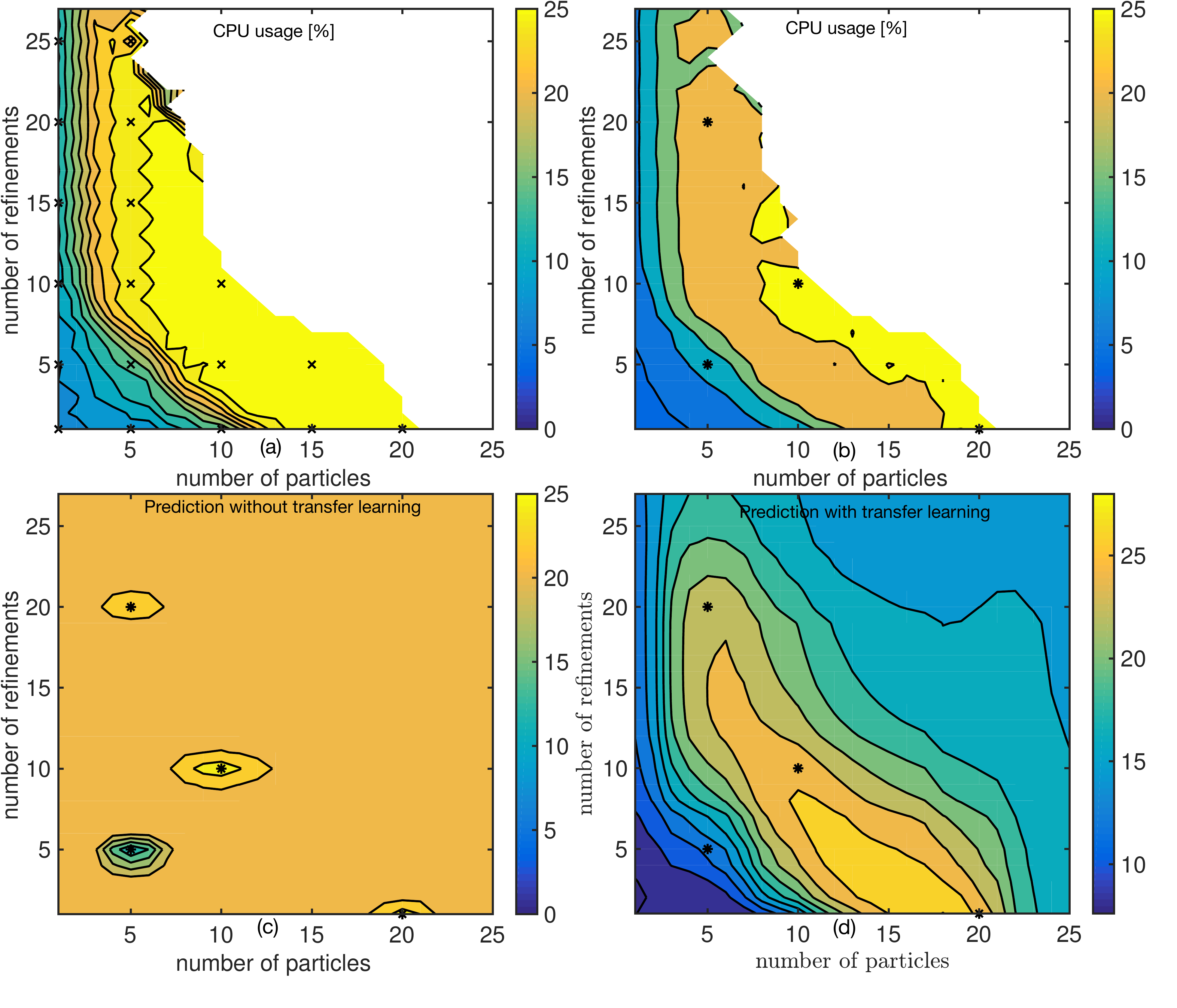}
		\caption{Illustration of transfer learning: (a) a source response function from which we can take samples; (b) a target response function which we are interested to learn; (c) a prediction without transfer learning; (d) a prediction with transfer learning.}
		\label{fig:example}
	\end{center}
\end{figure}

We have extensively measured the performance of the localization
component with different parameters and with different simulated
environmental conditions in prior work~\cite{kawthekarsensitivity}.
For the two parameters of interest (number of particles and refinements), we used 25 and 27 different values
each, evenly distributed on the log scale within their ranges ($[5-10,000]$ and $[1-10,000]$),
\emph{i.e.}, $|\mathbb{X}|=25\times 27=675$.
Specifically, we have repeatedly executed a specific mission to navigate along
a corridor in the simulator and measured performance in terms of CPU
usage (a proxy for energy consumption), time to completion, and location accuracy.
Each measurement takes about 30 seconds. In Figure~\ref{fig:example}a, we illustrate how CPU usage depends on those parameters
(the white area represents configurations in which the mission fails to complete).

As a transfer scenario, we consider simulation results given
different environment conditions.
As a target, we consider measurements that we have performed in the
default configuration.
As a source, we consider more challenging environment conditions,
in which the odometry sensor of the robot is less reliable
(\emph{e.g.}, due to an unfamiliar or slippery surface); we simulate
that the odometry sensor is both miscalibrated ($30\,\%$) and noisy ($\pm45\,\%$). We assume that we have collected these measurements
in the past or offline in a simulator and that those measurements
were therefore relatively cheap.
As shown in Figure~\ref{fig:example}b, localization becomes
more computationally expensive in the target environment.
Our goal is to use (already available) data from the noisy environment for
learning predictions in the non-noisy target environment with only a few
measurements in that environment.

%We have empirically observed that the cost of taking samples is approximately 40 times higher than simulator samples. Let us now assume for obtaining a source (target) sample the following cost incurred $c_s=0.3,c_t=12$ and we consider an experimental budget of $\mathcal{C}_{max}=250$.

\paragraph*{Observations}
For both source and target environments, we can learn fairly accurate
models if we take large numbers of samples (say $|\mathcal{D}|>80\,\%\times |\mathbb{X}|$) in that environment, but predictions
are poor when we can sample only a few configurations (say $|\mathcal{D}|<1\,\%\times |\mathbb{X}|$).
That is, we need at least a moderate number of measurements to
produce useful models for making self-adaptation decisions at runtime.
Also, using the model learned exclusively from measurements in the
source environment to predict the performance of the robot in the
more noisy environment leads to a significant prediction error
throughout the entire configuration space (cf. Figure \ref{fig:example}c).
However, transfer learning can effectively calibrate a model
learned from cheap measurements in the source environment to the
target environment with only a few additional measurements
from that environment. For example, Figure \ref{fig:example}d shows the predictions provided by a model that has been learned by transfer learning using 18 samples from the source and only 4 samples from the target environment. As a result, the model learned the structure of the response properly in terms of changing of values over the configuration space.

\begin{figure}[t]
	\begin{center}
		\includegraphics[width=\columnwidth]{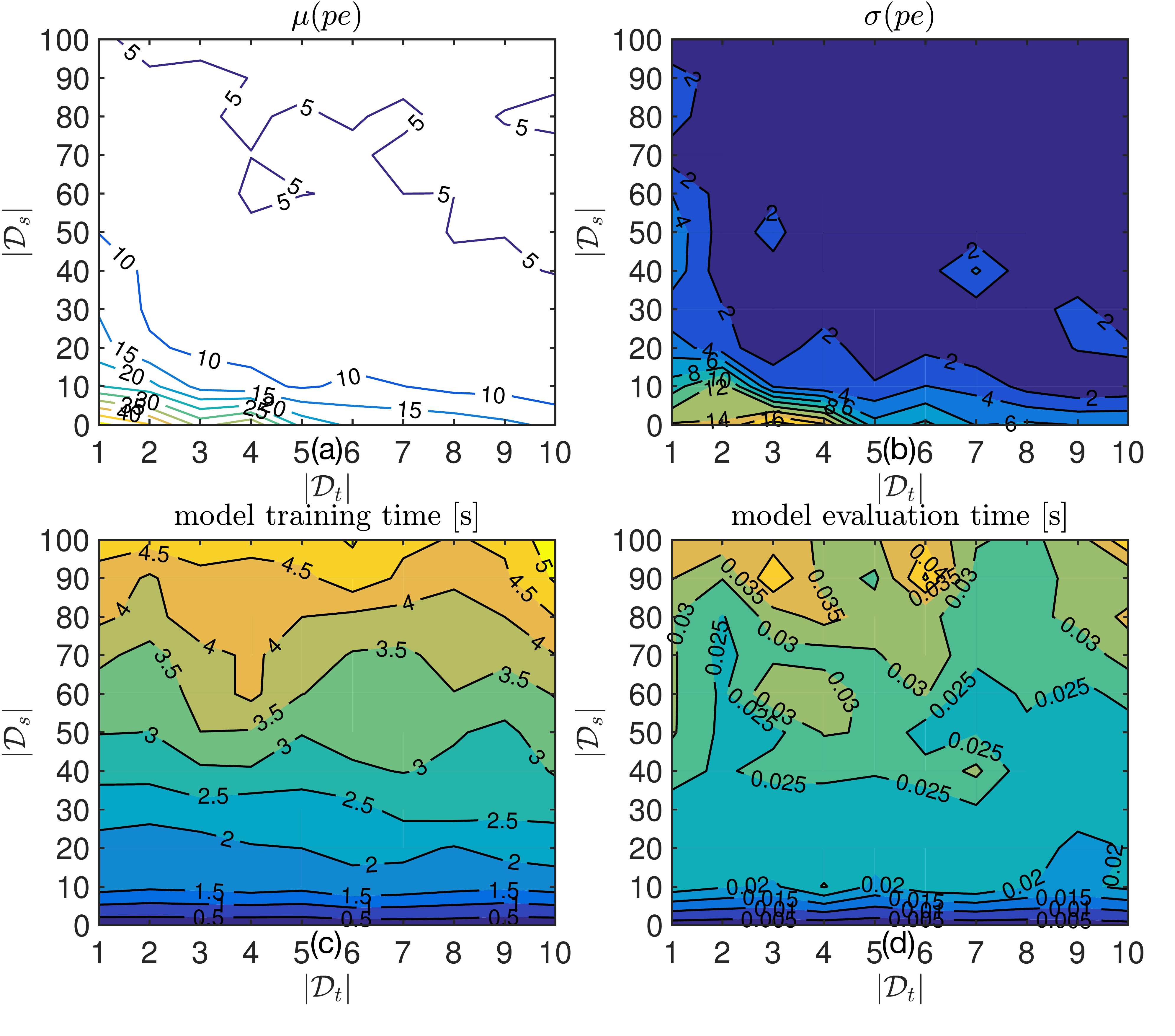}
		\caption{(a) Prediction error of the trained model, (b) variance of predictions (transfer learning contributes to lower the uncertainty regarding model predictions, therefore, more reliable learning), (c) training and (d) evaluation overhead.}
		\label{fig:pareto-cobot-mean-variance}
	\end{center}
\end{figure}

In Figure \ref{fig:pareto-cobot-mean-variance}a, we illustrate how well we can predict the target
environment with a different number of measurements in the source
and the target environment. More specifically, we randomly sampled configurations from both source and target between $[0-100\%]$ and $[0-10\%]$ respectively and we measured the accuracy as the relative error comparing predicted to actual performance in all 675 configurations. Note $|\mathcal{D}_s| = 0$ corresponds to model learning without transfer learning.
%We illustrated average and variance of prediction error over 3 repetitions with random samples in both environments in Figure \ref{fig:pareto-cobot-mean-variance}a,b respectively.
We can see that, in this case, predictions can be similarly accurate when we replace
expensive measurements from the target environment
by more, but cheaper samples.
Moreover, we have noticed that the models learned without transfer learning are highly sensitive to the sample selection (cf. Figure \ref{fig:pareto-cobot-mean-variance}b). The variance becomes smaller when we increase the number of source samples (along with the y-axis) leading to more reliable models.
In this case study, all model learning and evaluation times (both for normal learning and transfer learning) are negligible ($<5s$), cf. Figure \ref{fig:pareto-cobot-mean-variance}c,d.

\begin{figure}[t]
	\begin{center}
		\includegraphics[width=0.7\columnwidth]{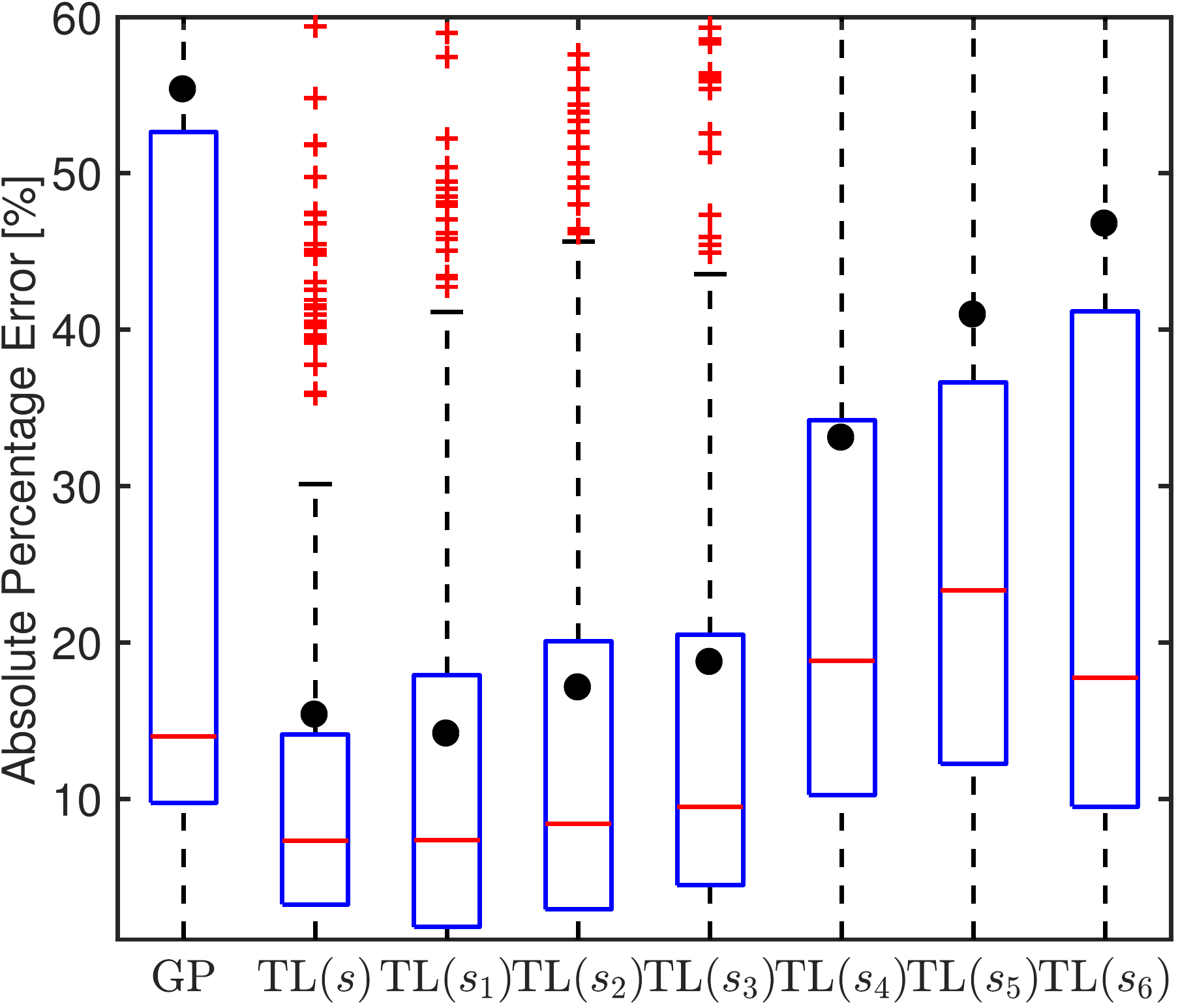}
		%\begin{table}{Example Table (floating)}
			%\caption{Sources with different levels of relatedness.}
			\resizebox{\columnwidth}{!}{%
				\begin{threeparttable}
					\begin{tabular}{lccccccc}
						\toprule
						Sources & $s$ & $s_1$ & $s_2$ & $s_3$ & $s_4$ & $s_5$ & $s_6$\\
						\midrule
						noise-level & $0$ & $5$ & $10$ & $15$ & $20$ & $25$ & $30$ \\
						corr. coeff. & $0.98$ & $0.95$ & $0.89$ & $0.75$ & $0.54$ & $0.34$ & $0.19$ \\
						$\mu(pe)$ & $15.34$ & $14.14$ & $17.09$ & $18.71$ & $33.06$ & $40.93$ & $46.75$ \\
						\bottomrule
					\end{tabular}
				\end{threeparttable}}
				\caption{Prediction accuracy of the model learned with samples from different sources of different relatedness to the target. GP is the model without transfer learning.}
				\label{fig:sensitivity_corr_boxplot}
			%\end{table}
	\end{center}
\end{figure}

\paragraph*{Relatedness}
In this case study, we can also observe that the relatedness between the source and target matters by changing the environments more or less aggressively.
For that purpose, we simulated six alternative source environments by adding noise with different power levels to the source we considered previously (\emph{i.e.}, odometry miscalibration of $30\,\%$ and noise of $\pm45\,\%$, cf., Figure~\ref{fig:sensitivity_corr_boxplot}). Typically,
the stronger the target environment is changed from the source (default) environment, the larger
the difference in performance is observed for the same configuration between the
two environments. This relationship is visible in correlation measures in
Figure~\ref{fig:sensitivity_corr_boxplot}). The prediction error reported in Figure \ref{fig:sensitivity_corr_boxplot} indicates that transfer learning becomes more effective
(\emph{i.e.}, produces accurate models with a few samples from the target environment)
if measurements in the source and target are strongly correlated. This also demonstrates that even transfer from a response with a small correlation helps to learn a better model comparing with no transfer learning.

Based on this observation, we need to decide (i) from which alternative source and (ii) how many samples we transfer to derive an accurate model. In general, we can: either (a) select the most related source and sample only from that source, or (b) select fewer samples from unrelated sources and more from the more related ones to transfer.
Note that our transfer learning method supports both strategies. For this purpose, we use the concept of relatedness (correlation) between source and target. The former is simpler, but we need to know, a priori, the level of relatedness of different sources. The latter is more sophisticated using a probabilistic interpretation of relatedness to learn from different sources using a concept of distance from the target. Our method supports this through the kernel function in Eq. \ref{eq:mt-kernel} that embeds the knowledge of source correlations and exploits that knowledge in the model learning process. The selection of a specific strategy depends on several factors including: (i) the cost of samples from each specific source (samples from some sources may come for free, \emph{i.e.} $c_s=0$ and we may want to use more samples from this source despite the fact that it may be less relevant), (ii) the bound on runtime overhead (more samples leads to a higher learning overhead), and (iii) domain for which the model is used (some domain are critical and some adversaries may manipulate the input data exploiting specific vulnerabilities of model learning to compromise the whole system security \cite{papernot2016towards}). We leave the investigation of strategy selection as a future work.

\subsection{Experiment: Prediction accuracy and model reliability}

With our case study, we demonstrated our approach on a specific small example.
To increase both internal and external validity, we perform a more systematic
exploration on multiple different configurable systems.

\paragraph*{Subject systems}
First, we again use the localization component of the CoBot system, but (since we are no longer constrained by plotting results in two dimensions) explore a larger space with 4 parameters.
In addition, we selected highly-configurable systems as subjects that have been explored for
configuration optimization in prior work~\cite{jamshidi2016bo4co}: three stream processing applications on Apache Storm
({\sf \small WordCount, RollingSort, SOL}) and a NoSQL database benchmark system on Apache Cassandra.
{\sf \small WordCount} is a popular benchmark~\cite{ghazal2013bigbench} featuring a three-layer architecture that counts the number of words in the incoming stream and it is essentially a CPU intensive application. {\sf \small RollingSort}  is a memory intensive system that performs rolling counts of incoming messages for identifying trending topics.
{\sf \small SOL} is a network intensive system, where the incoming messages will be routed through a multi-layer network.
%These are standard benchmarks that are widely used in the research community~\cite{ghazal2013bigbench} as well as industry benchmarks~\cite{huang2010hibench}. 
From the practical perspective, these benchmark applications are based on popular big data engines (\emph{e.g.}, Storm) and understanding the performance influence of configuration parameters can have a large impact on the maintenance of modern systems that are based on these highly-configurable engines. The Cassandra measurements was done using scripts that runs YCSB \cite{cooper2010benchmarking} and was originally developed for a prior work \cite{artavc2017dice}.

The notion of source and target set depends on the subject system.
In CoBot, we again simulate the same navigation mission in the default environment (source) and
in a more difficult noisy environment (target).
For the three stream processing applications, source and target
represent different workloads, such that we transfer measurements
from one workload for learning a model for another workload. More specifically, we control the workload using the maximum number of messages which we allow to enter the stream processing architecture.
For the NoSQL application, we analyze two different transfers:
First, we use as a source a query on
a database with 10 million records and as target the same query
on a database with 20 million records, representing a more expensive environment to sample from. Second, we use as a source a query on 20 million records on one cluster and as target a query on the same
dataset run on a different cluster, representing hardware changes.
Overall, our subjects cover different kinds of applications and
different kinds of \emph{transfer scenarios} (changes in the environment,
changes in the workload, changes in the dataset, and changes
in the hardware).

% the changes are along two different directions.
% First, we created a database of two different sizes: in the first, the database was loaded with 10 million records ({\sf cass-10}), while in the second it was loaded with 20 million records ({\sf cass-20}). We have also deployed each of these two versions of the benchmark on two different infrastructure setting leading to a total of 4 different response functions. %Figure~\ref{fig:cass-surf-all} (a,b,c,d) shows the response surfaces for the 4 versions of the NoSQL database where {\sf concurrent\_read} and {\sf concurrent\_write}, as configuration parameters, are varied ({\sf cass-10,20} are of different sizes, while $v_1,v_2$ are different in terms of deployment settings).

\paragraph*{Experimental setup}
As independent variables, we systematically vary the size of the
learning sets from both source and target environment in each
subject system.
We sample between 0 and~100\,\% of all configurations in the source
environment and between 1 and~10\,\% in the
target.

As a dependent variable, we measure learning time and prediction accuracy of the learned model.
For each subject system, we measure a large number of random configurations
as the evaluation set, independently from configurations sampled for learning,
and compare the predictions
of the learned model $\hat{f}$ to the actual measurements of the configurations in the evaluation
set $\mathcal{D}_o$. We compute the \emph{absolute percentage error (APE)}
for each configuration $x$ in the evaluation set $\frac{|\hat{f}(\mathbf{x})-f(\mathbf{x})|}{f(\mathbf{x})}\times 100$
and report the average to characterize the accuracy of the prediction model.
% and the \emph{normalized mean square error (NMSE)} $\frac{\Sigma(\hat{f}(\mathbf{x})-f(\mathbf{x}))^2}{|\mathcal{D}_o|.\sigma(\hat{\mathbf{y}})}$.
% The second accuracy metric is more sensitive to large deviations, i.e., it penalizes models that gives occasional large errors.
Ideally, we would use the whole configuration space as evaluation set ($\mathcal{D}_o=\mathbb{X}$), but the measurement effort would be prohibitively high for most real-world systems~\cite{jamshidi2016bo4co,influence}; hence we use large random samples (cf. size column in Table \ref{tab:configuration-parameters}).

The measured and predicted metric depends on the subject system:
For the CoBot system, we measure average CPU usage during the same mission of navigating along a corridor as in our case study;
we use the average of three simulation runs for each configuration.
For the Storm and NoSQL experiments, we measure average latency over a window of 8 and 10 minutes respectively.
Also, after each sample collection, the experimental testbed was cleaned which required several minutes for the Storm measurements and around 1 hour (for offloading and cleaning the database) for the Cassandra measurements.
We sample a given number of configurations in the source and target
randomly and report average results and standard deviations
of accuracy and learning time over 3 repetitions.

%\begin{figure*}[t]
%	\begin{center}
%		\includegraphics[width=\textwidth]{figures/cass-surfaces-all}
%		\caption{Response functions corresponding to the 2D subspace of {\sf cass-10,20} that are different in terms of \emph{infrastructure}.}
%		\label{fig:cass-surf-all}
%	\end{center}
%\end{figure*}

\begin{table}[t]
	\centering
	\caption{Overview of our experimental datasets. ``Size'' column indicates the number of measurements in the datasets and ``Testbed'' refer to the infrastructure where the measurements are taken and their details are in the appendix.} %Note $pe_{GP},pe_{TL}$ are the average APE resulted by the models trained without and with transfer learning.}
	\label{tab:configuration-parameters}
	\resizebox{0.8\columnwidth}{!}{%
		\begin{threeparttable}
			\begin{tabular}{@{}lllgc@{}}
				\toprule
				&\textbf{Dataset} & \multicolumn{1}{c}{\textbf{Parameters}}                                                                                       & \multicolumn{1}{c}{\textbf{Size}} & \multicolumn{1}{c}{\textbf{Testbed}} %  & \multicolumn{1}{c}{{$pe_{GP}$}} & \multicolumn{1}{c}{{$pe_{TL}$}}
				    \\ \midrule
				1  & {\sf CoBot(4D) }               & \begin{tabular}[c]{@{}l@{}}{\sf 1-odom\_miscalibration,}\\{\sf 2-odom\_noise,}\\{\sf 3-num\_particles,}\\ {\sf 4-num\_refinement}\\ \end{tabular}                      & 56585 & C9  % & 77.12 & 25.74
				\\
				\midrule
				2  & {\sf wc(6D) }               & \begin{tabular}[c]{@{}l@{}}{\sf 1-spouts, 2-max\_spout, }\\ {\sf 3-spout\_wait, 4-splitters,}\\ {\sf 5-counters, 6-netty\_min\_wait} \end{tabular}                      & 2880 & C1   % & 77.45&  0.44
				\\ \midrule
				3  & {\sf sol(6D) }             & \begin{tabular}[c]{@{}l@{}}{\sf 1-spouts, 2-max\_spout,} \\ {\sf 3-top\_level, 4-netty\_min\_wait,} \\ {\sf 5-message\_size, 6-bolts} \end{tabular}                          & 2866 & C2      %   & 89.12 & 0.11
				   \\ \midrule
				4  & {\sf rs(6D) }           & \begin{tabular}[c]{@{}l@{}}{\sf 1-spouts, 2-max\_spout, }\\ {\sf 3-sorters, 4-emit\_freq,}\\ {\sf 5-chunk\_size, 6-message\_size} \end{tabular}                                    & 3840 & C3   %    & 103.12 & 0.21
				  \\ \midrule
				\begin{tabular}[c]{@{}l@{}}5 \\ \\ 6\end{tabular} & \begin{tabular}[c]{@{}l@{}} {\sf cass-10 } \\ \\ {\sf cass-20 }\end{tabular}      & \begin{tabular}[c]{@{}l@{}}{\sf 1-trickle\_fsync, 2-auto\_snapshot,} \\ {\sf 3-con.\_reads, 4-con.\_writes} \\ {\sf 5-file\_cache\_size\_in\_MB}\\ {\sf 6-con.\_compactors} \end{tabular}     & 1024 & C6x,C6y \\ %&56.45 & 0.56  \\
				\bottomrule
			\end{tabular}
		\end{threeparttable}}
	\end{table}

\begin{figure*}[t]
	\begin{center}
		\includegraphics[width=0.95\textwidth]{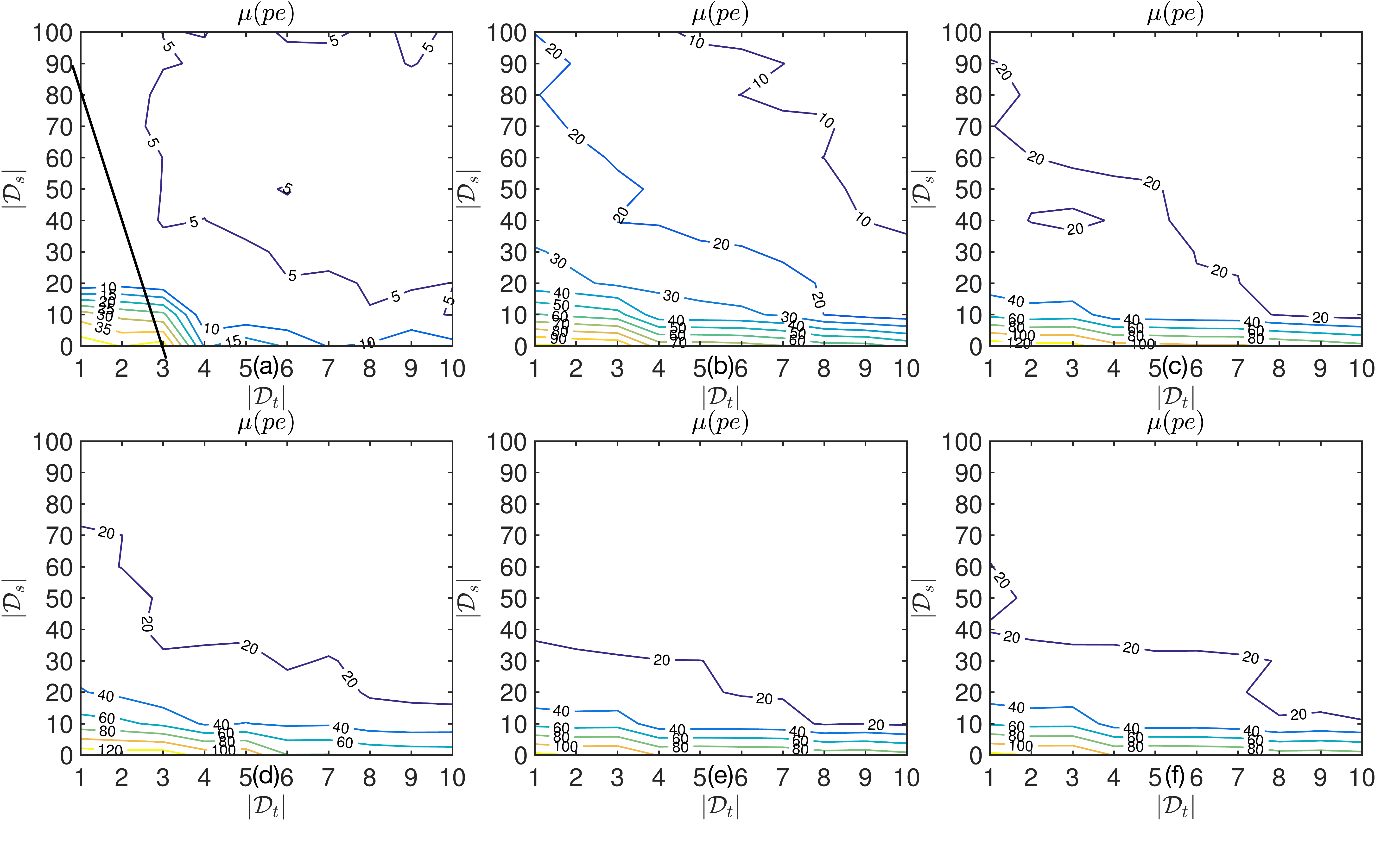}
		\caption{Prediction accuracy for (a) CoBot, (b) {\sf \small WC}, (c) {\sf \small SOL}, (d) {\sf \small RS}, (e) {\sf \small cass} (hardware change), (f) {\sf \small cass} (DB size change).}
		\label{fig:errors2dallsystems}
	\end{center}
\end{figure*}

\paragraph*{Results}
We show results of our experiments in Figure~\ref{fig:errors2dallsystems}. The 2D plot shows average errors across all subject systems.
The results in which the set of source samples $\mathcal{D}_s$ is empty represents the baseline case without transfer learning.
%In Figure~\ref{fig:accuracy-all}, we additionally show a specific slice through our accuracy results, in which we only vary the number of samples from the source (and only for 4 subject systems to produce a reasonably clear plot), but keep the number of samples from the target at a constant 1\,\%.
Although the results differ significantly among subject systems (not surprising, given different relatedness of the source and target) the overall trends are consistent.

% \begin{figure}[h]
% 	\begin{center}
% 		\includegraphics[width=0.7\columnwidth]{figures/aveErr_perc_source}
% 		\caption{Average prediction error for the subject systems with a fixed target sample size.}
% 		\label{fig:accuracy-all}
% 	\end{center}
% \end{figure}

First, our results show that transfer learning can achieve high prediction accuracy with only a few samples from the target environment;
that is, transfer learning is clearly beneficial if samples from the source environment are much cheaper than samples from the target environment.
Given the same number of target samples, the prediction accuracy becomes better up to even 3 levels of magnitudes as we include more samples from the source.

Second, using transfer learning with larger samples from the source environment reduces  the variance of the prediction error. That is, models learned with only a few samples from the target environment are more affected by the randomness in sampling from the configuration space. Conversely, transfer learning improves the quality and
reliability of the learned models, by reducing sensitivity to the sample selection.

Third, the model training time increases monotonically when we increase the sample size for target or source (cf. Figure \ref{fig:pareto-cobot-mean-variance}c for the CoBot case study and for the other subject systems in the appendix). At the same time,
even for large samples, it rarely exceeds 100 seconds for our subject systems, which is reasonable for online use in a self-adaptive system. Learning time is negligible compared to measurement times in all our subject systems.

Finally, the time for evaluating the models (\emph{i.e.}, using
the model to make a prediction for a given configuration), does not change
much with models learned from different sample sizes (cf. Figure \ref{fig:pareto-cobot-mean-variance}d) and is negligible overall ($<300$ms).

\begin{shaded}
Summary: we see improved accuracy and reliability
when using transfer learning with additional source samples (RQ1); we
see clear trade-offs among using more source or target samples (RQ2);
and we see that training and evaluation times are acceptable to be applied for self-adaptive systems (RQ3).
\end{shaded}

%\begin{figure}
%	\begin{center}
%		\includegraphics[width=0.7\columnwidth]{figures/training_time_cobot}
%		\caption{Model training time with different source and target samples for the CoBot experiment}
%		\label{fig:training-testing-runtime}
%	\end{center}
%\end{figure}

% \begin{figure}
% 	\begin{center}
% 		\includegraphics[width=\columnwidth]{figures/runtime_all}
% 		\caption{Model learning and evaluation time.}
% 		\label{fig:training-testing-runtime-all}
% 	\end{center}
% \end{figure}

%\begin{figure}
%	\begin{center}
%		\includegraphics[width=0.7\columnwidth]{figures/pareto-sps-annotated1}
%		\caption{Indifference curves for an stream processing system.}
%		\label{fig:pareto-sps}
%	\end{center}
%\end{figure}

%\begin{figure}
%	\begin{center}
%		\includegraphics[width=0.7\columnwidth]{figures/model_reliability}
%		\caption{Variance of prediction error with and without transfer learning and different number of sample sets.}
%		\label{fig:model-reliability}
%	\end{center}
%\end{figure}

\subsection{Discussion: Trade-offs and cost models}
Our experimental results clearly show that we can achieve
accurate models both with a larger number of samples from source
or target environment or both:
In Figure~\ref{fig:errors2dallsystems}, we can see how
many models for different numbers of source and target samples
yield models with similar accuracy.
Those different points with the same accuracy can be interpreted
as indifference curves (a common tool in economics~\cite{binger1988microeconomics}).
We can invest different measurement costs to achieve models with
equivalent utility.
In Figure~\ref{fig:errors2dallsystems}a,
we show those indifference curves more explicitly for the CoBot
system. Using the indifference curves enables us to decide how many samples from either a source or a target we should use for the model learning process that is most beneficial in increasing predictive accuracy over unseen regions of the configuration space while satisfying the budget constraint.
Given concrete costs and budgets, say, samples from the target environment are 3 times
more expensive to gather than samples from the source environment,
we plotted an additional line representing our budget and find the
combination of source and target samples that produces the best
prediction model for our budget (see intersection of the budget
line with the highest indifference curve in Figure~\ref{fig:errors2dallsystems}).
Although we will not know the indifference lines until we run
extensive experiments, we expect that the general trade-offs
are similar across many subject systems and transfer scenarios
and that such a cost model (cf. Eq. \ref{eq:cost-model}) can help to justify decisions
on how many samples to take from each environment.
We leave a more detailed treatment to future work.

\begin{figure}[t]
 	\begin{center}
 		\includegraphics[width=0.8\columnwidth]{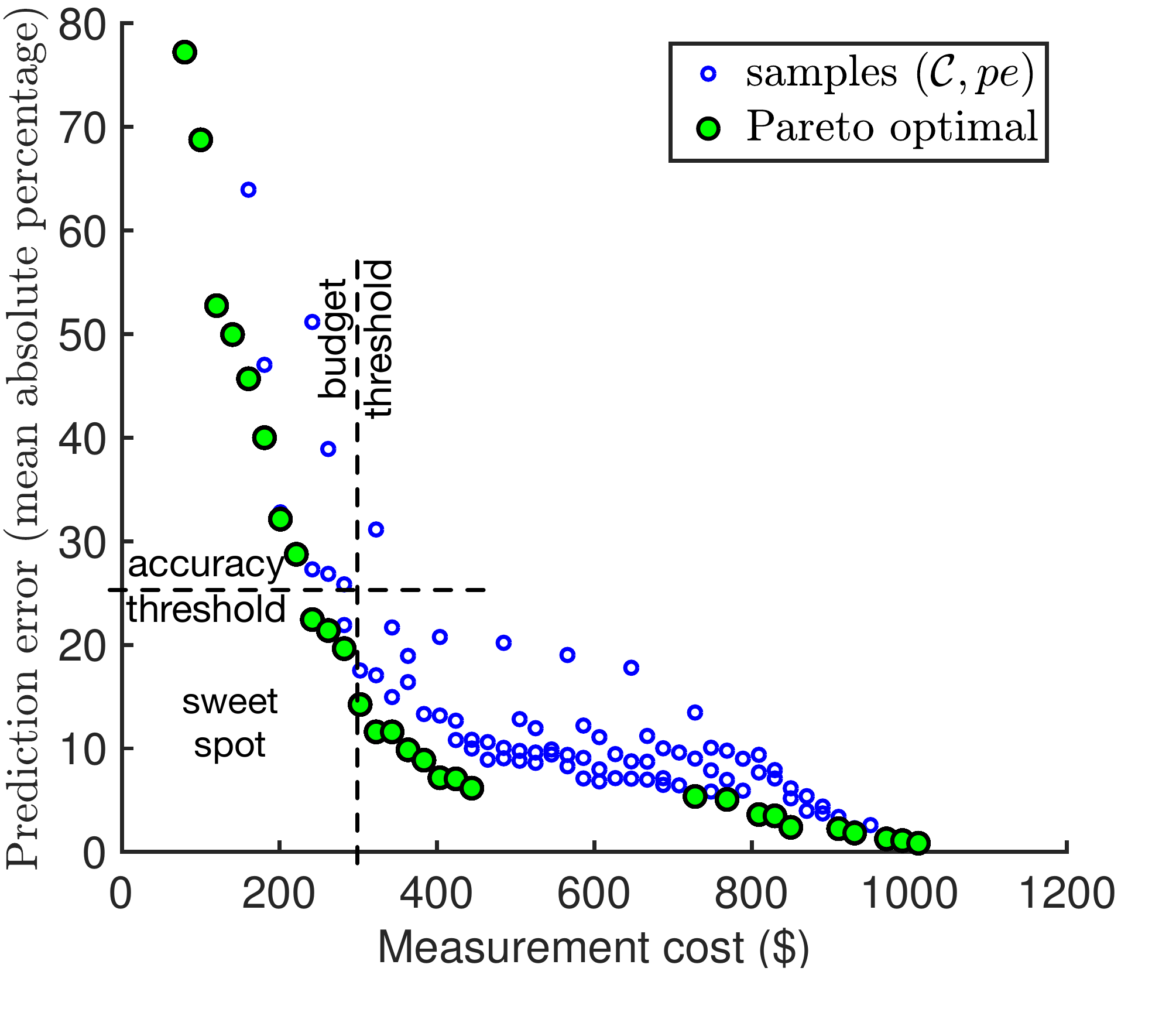}
 		\caption{A two-objective optimization goal. We are interested in the solutions in the targeted sweet spot.}
 		\label{fig:pareto-cobot-effort-accuracy}
 	\end{center}
 \end{figure}

In this context, we can fix a cost model (\emph{i.e.}, fixing specific values for $c_s,c_t$) and transform the indifference diagrams into a multi-objective goal and derive the Pareto front solutions as shown in Figure \ref{fig:pareto-cobot-effort-accuracy}. This helps us to locate the feasible Pareto front solutions within the sweet spot.

\subsection{Threats to validity and limitations}

%{\noindent \em Internal validity.}
\subsubsection{Internal validity}
In order to ensure internal validity, we repeated the execution of the benchmark systems and measure performance for a large number of configurations. %We repeated this at least 3 times for each configuration to avoid measurement bias.
For doing so, we have invested several months for gathering the measurements, which resulted in a substantial dataset. Moreover, we used standard benchmarks so that we are confident in that we have measured a realistic scenario.

\subsubsection{External validity}
%{\noindent \em External validity.}
In order to ensure external validity, we use three classes of systems in our experiments including: (i)~a robotic system, (ii)~3 different stream processing applications, and (iii)~a NoSQL database system. These systems have different numbers of configuration parameters and are from different application domains.

\subsubsection{Limitations}
%{\noindent \em Limitations.}
Our learning approach relies on several assumptions. First, we assume that the target response function is smooth. If a configuration parameter has an unsteady performance behavior, we cannot learn a reliable model, but only approximate its performance close to the observations. Furthermore, we assume that the source and target responses are related. That is they are correlated to a certain extent. The more related, the faster and better we can learn. Also, we need the configurable system to have a deterministic performance behavior. If we replicate the performance measurements for the same system, the observed performance should be similar. %Finally, our approach, because of GP limitations, has its limits regarding the number of configuration parameters and the size of the learning set. For instance, it is infeasible to learn a model with thousands of configuration parameters. However, we could combine our approach with a dimensionality reduction technique to determine and learn only for the relevant parameters, as this have been shown before~\cite{influence}.

\newpage
%\input{sections/discussion}
%\input{sections/relatedwork}
%!TEX root = ../paper.tex
\section{Conclusions}
\label{sec:conclusions}

Today most software systems are configurable and performance reasoning is typically used to adapt the configuration in order to respond to environmental changes. Machine learning and sampling techniques have been previously proposed to build models in order to predict the performance of unseen configurations. However, the models are either expensive to learn, or they become extremely unreliable if trained on sparse samples. Our cost-aware transfer learning method is orthogonal to the both previously proposed directions promoting to learn from other cheaper sources. Our approach requires only very few samples from the target response function and can learn an accurate and reliable model based on sampling from other relevant sources. We have done extensive experiments with 5 highly configurable systems demonstrating that our approach (i) improves the model accuracy up to several orders of magnitude, (ii) is able to trade-off between different number of samples from source and target, and (iii) imposes an acceptable model building and evaluation cost making appropriate for application in the self-adaptive community.  

{\noindent \em Future directions.}
Beside performance reasoning, our cost-aware transfer learning can contribute to reason about other non-functional properties and quality attributes, \emph{e.g.}, energy consumption. Also, our approach could be extended to support configuration optimization. For example, in our previous work, we have used GP models with Bayesian optimization to focus on interesting zones of the response functions to find optimum configuration quickly for big data systems \cite{jamshidi2016bo4co}. %We found that, if we learn from an unrelated source, it leads to a negative transfer and the predictions become worse comparing with no transfer learning. Hence, it is important to choose a relevant source for model learning.
%In general, our notion of cost-aware transfer learning is independent of the concrete learning technique. In this work, we have used GP models, because our experimental results indicated that GP models outperform other regression models for the systems under consideration \cite{jamshidi2016bo4co}. 
In general, our notion of cost-aware transfer learning is independent of a particular black-box model and complementary to white-box approaches in performance modeling such as queuing theory. A more intelligent way (\emph{e.g.}, active learning) of sampling the source and the target environment to gain more information is also another fruitful future avenue.

\section*{Acknowledgment}
This work has been supported by AFRL and DARPA (FA8750-16-2-0042). Kaestner's work is also supported by NSF awards 1318808 and 1552944 and the Science of Security Lablet (H9823014C0140). Siegmund's work is supported by the DFG under the contract SI 2171/2.

\newpage

%\vfill
\bibliographystyle{abbrv}
%\bibliography{sigproc}  % sigproc.bib is the name of the Bibliography in this case
{\footnotesize
\bibliography{paper.bib}}
%\appendix

%\input{sections/appendix}

\end{document}